# Space-Based Coronagraphic Imaging Polarimetry of the TW Hydrae Disk: Shedding New Light on Self-Shadowing Effects

Charles A. Poteet,[1] Christine H. Chen,[1,2] Dean C. Hines,[1] Marshall D. Perrin,[1] John H. Debes,[1] Laurent Pueyo,[1] Glenn Schneider,[3] Johan Mazoyer,[2,1] and Ludmilla Kolokolova[4]

[1]Space Telescope Science Institute, 3700 San Martin Drive, Baltimore, MD, 21218, USA; charles.poteet@gmail.com
[2]Department of Physics and Astronomy, Johns Hopkins University, Baltimore, MD 21218, USA
[3]Steward Observatory, The University of Arizona, 933 North Cherry Avenue, Tucson, AZ 85721, USA
[4]Department of Astronomy, University of Maryland, College Park, MD 20742, USA

## ABSTRACT

We present *Hubble Space Telescope* Near-Infrared Camera and Multi-Object Spectrometer coronagraphic imaging polarimetry of the TW Hydrae protoplanetary disk. These observations simultaneously measure the total and polarized intensity, allowing direct measurement of the polarization fraction across the disk. In accord with the self-shadowing hypothesis recently proposed by Debes et al., we find that the total and polarized intensity of the disk exhibits strong azimuthal asymmetries at projected distances consistent with the previously reported bright and dark ring-shaped structures ($\sim$45–99 au). The sinusoidal-like variations possess a maximum brightness at position angles near $\sim$268–300° and are up to $\sim$28% stronger in total intensity. Furthermore, significant radial and azimuthal variations are also detected in the polarization fraction of the disk. In particular, we find that regions of lower polarization fraction are associated with annuli of increased surface brightness, suggesting that the relative proportion of multiple-to-single scattering is greater along the ring and gap structures. Moreover, we find strong ($\sim$20%) azimuthal variation in the polarization fraction along the shadowed region of the disk. Further investigation reveals that the azimuthal variation is not the result of disk flaring effects, but instead from a decrease in the relative contribution of multiple-to-single scattering within the shadowed region. Employing a two-layer scattering surface, we hypothesize that the diminished contribution in multiple scattering may result from shadowing by an inclined inner disk, which prevents direct stellar light from reaching the optically thick underlying surface component.

*Keywords:* circumstellar matter — protoplanetary disks — polarization — stars: individual (TW Hya) — stars: pre-main sequence

## 1. INTRODUCTION

Protoplanetary disks are revolving structures comprised of gas and dust that surround young stars. As a by-product of the star formation process, these disks are naturally considered to represent the birth sites of planets. High-resolution scattered light and thermal emission imaging of protoplanetary disks have revealed small-scale morphological structures that are associated with planet-disk interactions and disk evolution. Although direct evidence for the presence of planets has not been directly observed in the vast major of disk systems, insights into the process governing their formation is often inferred from the existence of central clearings (e.g., Avenhaus et al. 2014; Andrews et al. 2016), spiral arms (e.g., Garufi et al. 2013; Benisty et al. 2015; Follette et al. 2017), and gaps (e.g., Quanz et al. 2013; ALMA Partnership et al. 2015; Bae et al. 2017) in the surface and mid-plane layers of protoplanetary disks.

TW Hydrae (TW Hya, $K_s = 7.297$, $H - K_s = 0.261$; Cutri et al. 2003), located at a distance of 59.5 pc (Gaia Collaboration et al. 2016), is the nearest known example of a star (M0.5V; Herczeg & Hillenbrand 2014; Sokal et al. 2018) surrounded by a protoplanetary disk. Despite its relatively advanced age ($\sim$8 Myr; Donaldson et al. 2016), the nearly face-on disk ($i = 7 \pm 3°$; Qi et al. 2004; Rosenfeld et al. 2012; Andrews et al. 2016) possesses morphological structures indicative of on-going planet formation. Continuum observations with the Atacama Large Millimeter/submillimeter Array (ALMA) have revealed a series of concentric bright and dark rings at projected radii less than 50 au, including a gap centered at $\sim$1 au, a bright ring at $\sim$3 au (Andrews et al. 2016), and deep gaps near 24 and 41 au (Andrews et al. 2016; Tsukagoshi et al. 2016; Nomura et al. 2016). In addition, gaps at <7 au, ~22 au, and 88 au have also been detected in scattered light using ground-based coronagraphic imaging polarimetry (Akiyama et al. 2015; Rapson et al. 2015; van Boekel et al. 2017). Moreover, total scattered light observations with the Space Telescope Imaging Spectrograph (STIS) on the *Hubble Space Telescope* (*HST*) have revealed that the disk possesses an azimuthal surface brightness asymmetry that rotates in the counterclockwise direction at a constant angular velocity of $22°.7$ yr$^{-1}$ (Debes et al. 2017). The surface brightness asymmetry can be explained by the presence of a planet at 1.1 au, which induces a warp in the disk interior to 1 au and casts a shadow on the outer disk surface.

Shadows have also been detected in a number of other disk systems: HD 142527 (Avenhaus et al. 2014; Marino et al. 2015), LkCa 15 (Thalmann et al. 2014), HD 135344 B (Stolker et al. 2016), PDS 66 (Wolff et al. 2016), and HD 100453 A (Benisty et al. 2017; Long et al. 2017). The existence of shadows on the surface of protoplanetary disks may not only be the consequence of planet-induced warps, but may also be the product of a strongly misaligned inner disk with respect to its surrounding outer disk (e.g., Marino et al. 2015; Benisty et al. 2017). A misaligned inner disk may be caused by interactions with either the magnetic field of the central star or a massive companion on a highly inclined orbit (Min et al. 2017, and references therein). Moreover, shadows may also play an important role in the formation of spiral arm structures observed in protoplanetary disks (e.g., Montesinos et al. 2016; Benisty et al. 2017).

In this paper, we present *HST* Near-Infrared Camera and Multi-Object Spectrometer (NICMOS) coronagraphic imaging polarimetry of the TW Hya disk. The observations were obtained during the Cycle 12 commissioning of the NICMOS coronagraphic polarimetry mode and offer new insight into the scattered light morphological structure of the TW Hya disk compared with previously published ground-based polarimetric studies. Specifically, ground-based observations currently permit the measurement of the polarized scattered intensity of disks through the simultaneous measurement of the Stokes $Q$ and $U$ parameters. However, the polarized intensity must be interpreted in the context of the total intensity (Stokes $I$) to accurately constrain disk and dust grain properties (Perrin et al. 2009; Jang-Condell 2017). In general, ground-based total intensity imaging of disks requires post-processing techniques, such as Angular Differential Imaging methods (ADI; Marois et al. 2006), the Locally Optimized Combination of Images algorithm (LOCI; Lafrenière et al. 2007), the Karhunen-Loève Image Projection algorithm (KLIP; Soummer et al. 2012), or the Non-negative Matrix Factorization method (NMF; Ren et al. 2018), to recover the total intensity. However, these advanced techniques are not optimized for disks with face-on orientations, making estimates of the total intensity and polarization fraction challenging due to disk self-subtraction effects. On the contrary, space-based imaging polarimetry of face-on disks enables the combination of sequential observations with linear polarizers to simultaneously derive high quality Stokes $I$, $Q$, and $U$ parameters and thus the polarization fraction.

The organization of the paper is summarized as follows. In Section 2, we present an overview of the *HST*-NICMOS coronagraphic imaging polarimetry observations, data reduction, and image analysis. In Section 3, we describe the scattered light intensities and polarization fraction distributions of the TW Hya disk. In Section 4, we discuss the implications of our findings in the context of self-shadowing and multiple scattering effects. Finally, we summarize the main conclusions from our study in Section 5.

## 2. NICMOS CORONAGRAPHIC POLARIMETRY

### 2.1. *Observations*

Following standard observing practices for *HST* coronagraphy, TW Hya was observed with the NICMOS (Thompson et al. 1998) Camera 2 in four visits on 2004 February 2 and 2004 April 13. The observations were obtained as part of the Cycle 12 General Observer program 9768 (PI: D. Hines) and are summarized in Table 1. During each visit, two exposures (63.95 s and 79.94 s) were acquired in each of the long-wavelength linear polarizers: POL0L, POL120L, and POL240L ($\lambda = 2.05$ $\mu$m, $\Delta\lambda = 1.89$–$2.10$ $\mu$m). To maximize the dynamic range of the observations, the exposures were obtained using the STEP16 sampling sequence of the MULTIACCUM mode. The spacecraft orientation between consecutive visits was rotated by $29°.9$ to isolate rotationally invariant instrumental artifacts and minimize the effects of bad pixels.

Because the commissioning program was specifically designed to calibrate the coronagraphic polarimetry mode of NICMOS (Hines & Schneider 2006), it did not include dedicated point-spread function (PSF) reference observations. Consequently, suitable PSF reference stars were assembled from the Legacy Archive PSF Library And Circumstellar Environments program (LAPLACE; Schneider et al. 2010). However, due to the NICMOS Cooling System (NCS) failure in 2008, we find that only six PSF reference stars were observed using the coronagraphic polarimetry mode between Cycles 11 and 15: HD 21447, HIP 21556, HD 42807, CD $-75°514$, BD $+32°3739$, and GJ 784. These PSF reference candidates span a broad range in stellar spectral type (BV–MV), $H - K_s$ color, *HST* breathing phase (e.g., Bély et al. 1993), and NICMOS cold mask alignment (e.g., Krist et al. 1998). We note that CD $-75°514$ is a known polarimetric standard star and was excluded from further consideration.

Table 1. Summary of *HST*-NICMOS Coronagraphic Imaging Polarimetry Observations

| Visit Number | Observation Date | Spacecraft Orientation (°) | Exposure Time Sequence[a] (s) | Observation Set ID | Observation Number |
|---|---|---|---|---|---|
| TW Hya | | | | | |
| 30..... | 2004 Feb 02 | 88.57 | 63.95, 79.94 | 8QV30 | HP, HQ, HR, HS, HT, HU |
| 31..... | 2004 Feb 02 | 118.47 | 63.95, 79.94 | 8QV31 | HZ, I0, I1, I2, I3, I4 |
| 60..... | 2004 Apr 13 | 182.67 | 63.95, 79.94 | 8QV60 | FJ, FK, FL, FM, FN |
| 61..... | 2004 Apr 13 | 212.57 | 63.95, 79.94 | 8QV61 | FW, FX, FY, FZ, G0, G1 |
| BD +32°3739 (PSF Reference Star) | | | | | |
| 10..... | 2003 Nov 03 | 224.27 | 47.96, 47.96, 63.95 | 8QV10 | GG, GH, GI, GJ, GK, GL, GM, GO, GP |
| 11..... | 2003 Nov 03 | 194.37 | 47.96, 47.96, 63.95 | 8QV11 | GU, GV, GW, GX, GY, GZ, H0, H2, H3 |
| 40..... | 2004 Aug 02 | 314.57 | 47.96, 47.96, 47.96 | 8QV40 | FF, FG, FH, FI, FJ, FK, FL, FN, FO |
| 41..... | 2004 Aug 02 | 284.67 | 47.96, 47.96, 47.96 | 8QV41 | FS, FT, FU, FV, FW, FX, FY, G0, G1 |

NOTE—Observations from each visit were executed using the long-wavelength linear polarizers POL0L, POL120L, and POL240L, as part of Cycle 12 General Observer program 9768 (PI: D. Hines).

[a] Exposure time sequence per linear polarizer.

While all candidates satisfy the differential color criterion ($|\Delta(H - K_s)| < 0.3$) defined by Grady et al. (2010), preliminary PSF subtraction reveals significant mismatches in breathing phase and/or cold mask alignment when using HD 21447, HIP 21556, HD 42807, or GJ 784 as the PSF reference. Despite being a factor of ∼4 fainter in the $K_s$-band, we find that the remaining candidate, BD +32°3739 ($K_s = 8.721$, $H - K_s = 0.054$), is well-matched in breathing phase and cold mask alignment, and thus results in substantially better PSF subtraction. As a result of these findings, we adopt BD +32°3739 as the PSF reference star.

Coronagraphic observations of the unpolarized standard star BD +32°3739 (A6V; Schmidt et al. 1992) were similarly obtained during four visits on 2003 November 3 and 2004 August 2, as part of the Cycle 12 commissioning program. A summary of the observations are provided in Table 1. We note that observations from visit 10 are affected by thermally-induced changes in breathing phase and/or cold mask alignment and were excluded from this study.

## 2.2. *Data Reduction*

The instrumentally-calibrated and reduced *HST*-NICMOS coronagraphic images of TW Hya and BD +32°3739 were retrieved from the LAPLACE archive, which offers improvements over the standard calibration pipeline in dark subtraction, flat-field correction, and bad pixel correction. Given the low Galactic latitude ($b \simeq -0°.6$) of BD +32°3739, the PSF reference images were initially inspected for contamination by nearby ($\leqslant 3''.5$) field stars. Following the prescription of Lowrance et al. (1999), two foreground and/or background stars, located ∼$1''.4$ and ∼$2''.5$ from BD +32°3739, were identified by subtracting POL*L-specific images obtained at different spacecraft orientations. In order to minimize the impact of negative residuals in the PSF-subtracted images of the TW Hya disk, the point sources were modeled and removed from the PSF reference images using a two-dimensional Gaussian function.

Following an approach similar to that described in Pueyo et al. (2015), the precise location of each occulted star was measured using a Radon transform-based technique, which assumes that the diffraction spike pattern produced by the *HST* secondary mirror support structures is centered on the position of the occulted star. In this procedure, the speckle pattern surrounding the coronagraphic obscuration in each image was initially masked using a 40-pixel radius aperture. The images were then temporarily corrected for bad pixels, Laplacian-filtered, and additionally masked to retain only the diffraction spike pattern. The boundaries of each image were subsequently padded to ensure that the LAPLACE-estimated position of the occulted star was centered in the field-of-view. The Radon transform of the resulting images was then calculated to within a precision of 0.007 pixels in the $r$-direction and $0°.01$ in the $\theta$-direction. The polar coordinates of the brightest diffraction spikes were determined and projected as line segments in Cartesian space. Finally, the position of the occulted star relative to the center of each image

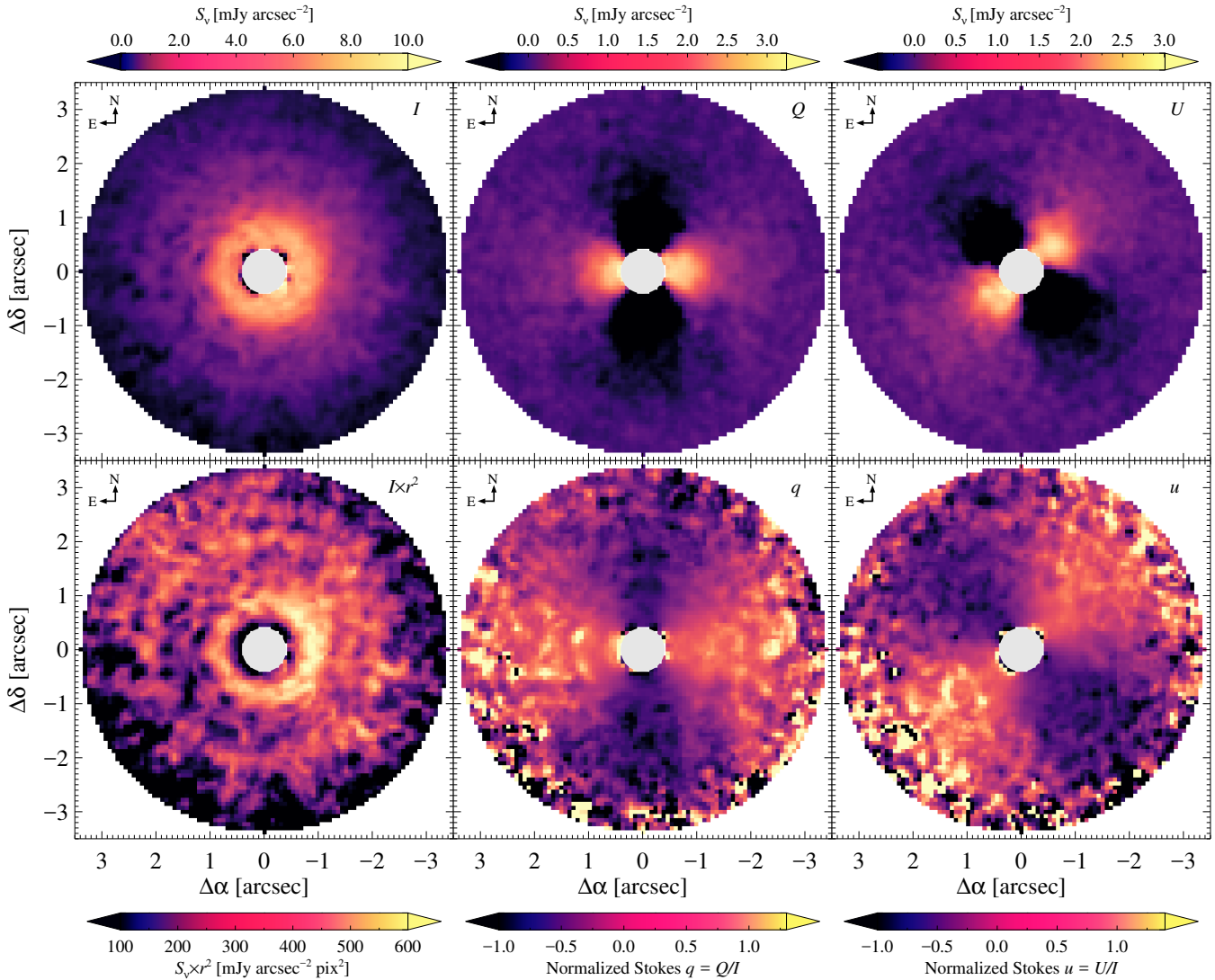

**Figure 1.** *HST*-NICMOS Stokes images of the TW Hya disk at 2.05 $\mu$m. Top panels: Stokes $I$ (total intensity; left), Stokes $Q$ (middle), and Stokes $U$ (right) images. Bottom panels: $r^2$-scaled Stokes $I$ (left), normalized Stokes $q$ (middle), and normalized Stokes $u$ (right) images. The $r^2$ factor represents the projected squared pixel distance from the central star, and is multiplied by the intensity to eliminate inverse-square illumination effects. All images have been smoothed using a one-resolution element ($\sim0\rlap.{''}22$) Gaussian, and are displayed using an inverse hyperbolic sine stretch. A $0\rlap.{''}4$-radius coronagraphic mask (light gray circle) is indicated at the center of each image.

was inferred from the intersection of the corresponding line segments.

Preceding image registration, dead and/or hot pixels in each image were identified using a $3\sigma$ filtering process and replaced by the mean value of the surrounding pixels. All images were aligned to a common reference frame using the occulted stellar positions derived from the Radon transform-based technique. Image alignment was performed to sub-pixel accuracy using a cubic convolution interpolation method (Park & Schowengerdt 1983). The average background of each image was subsequently measured using three 30×30 pixel regions located within the non-coronagraphic quadrants of the detector and subtracted to minimize the presence of DC bias offset. Finally, the POL*L-specific co-aligned images were median-combined for each spacecraft orientation.

We estimated the uncertainties associated with each POL*L-specific median-combined image assuming that the noise is uncorrelated and comprised of both statistical (shot noise, read noise, and dark current) and systematic sources (speckle noise). The systematic noise component was determined by differencing PSF reference images of BD +32°3739 at different epochs. In particular, we subtracted the POL*L-specific median-combined images obtained in visits 40 and 41 from that acquired in visit 11. To capture the variation in the systematic noise, we measured the standard deviation of the residual speckle pattern present in each image using concentric annuli at 3-pixel (∼$0''\!.23$) intervals from the occulted star. Finally, we averaged the resulting systematic uncertainty images and added the result in quadrature with the statistical uncertainty images of TW Hya and BD +32°3739 to produce POL*L-specific median-combined uncertainty images of each target.

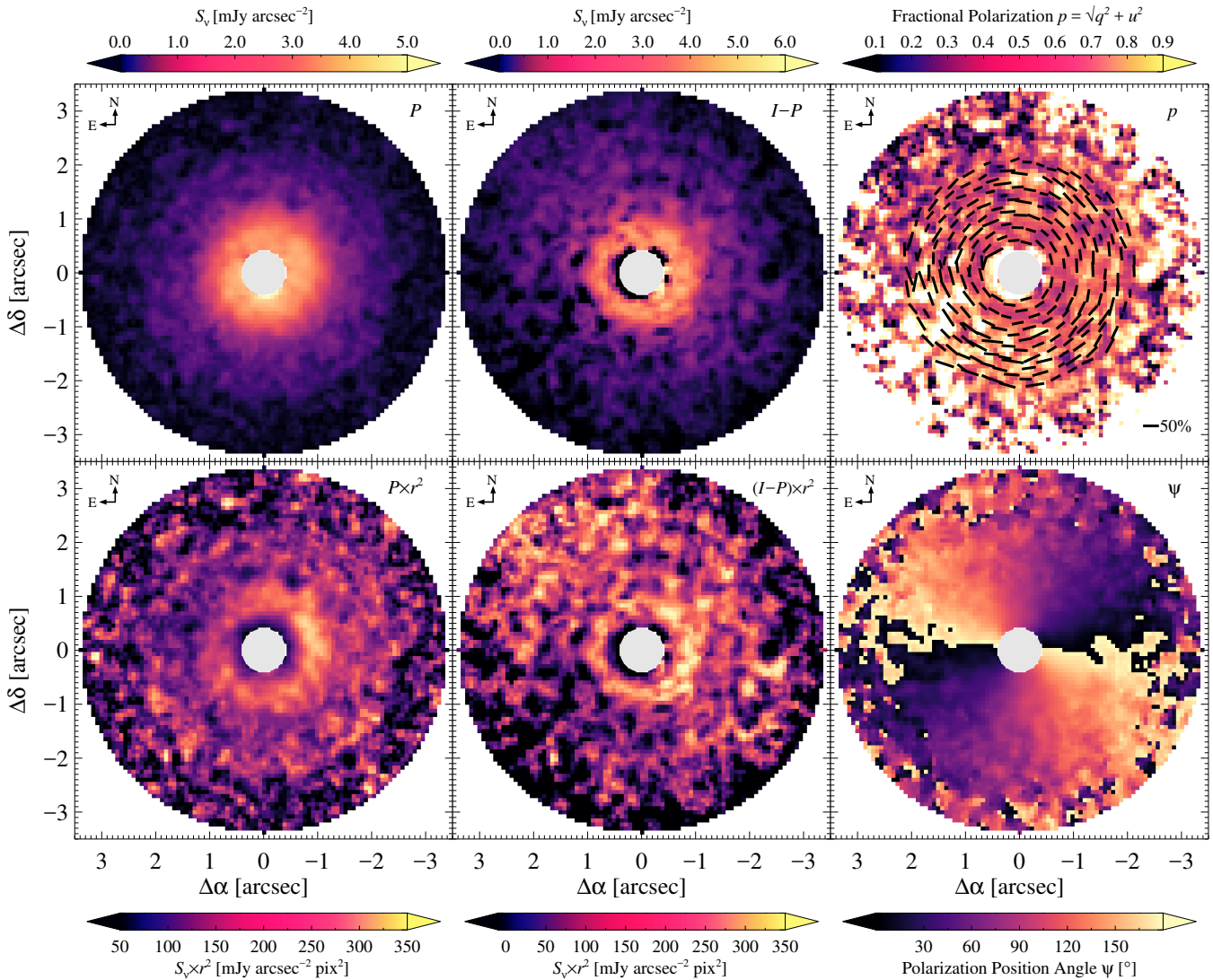

**Figure 2.** *HST*-NICMOS imaging polarimetry of the TW Hya disk at 2.05 μm. Top panels: polarized intensity (left), non-polarized intensity (middle), and fractional polarization (right) images. Bottom panels: $r^2$-scaled polarized intensity (left), $r^2$-scaled non-polarized intensity (middle), and polarization position angle (right) images. The $r^2$ factor represents the projected squared pixel distance from the central star, and is multiplied by the intensity to eliminate inverse-square illumination effects. The non-polarized intensity image was constructed by subtracting the polarized intensity from the total intensity. Polarization vectors (black lines) are illustrated for 10% of the data at projected radii within $2''\!.2$ of the central star. The polarization position angle image is displayed using a linear stretch, while the remaining images are displayed using an inverse hyperbolic sine stretch. All images have been smoothed using a one-resolution element (∼$0''\!.22$) Gaussian. A $0''\!.4$ radius coronagraphic mask (light gray circle) at the center of each image and missing data in the polarization fraction image (white pixels) are indicated.

### 2.3. *PSF Subtraction*

In order to reduce the instrumentally-defracted light in the coronagraphic images of TW Hya, we performed classical PSF subtraction using the unpolarized standard reference star BD +32°3739. After masking the 45-pixel radius region surrounding the coronagraphic obscuration, we determined the PSF scale factor between the POL*L-specific median-combined images of BD +32°3739 and TW Hya for each spacecraft orientation by minimizing the squared residual signal in the diffraction spike pattern. For the POL0L, POL120L, and POL240L polarizers, we found the average PSF scale factors to be $3.398 \pm 0.068$, $3.607 \pm 0.025$, $3.341 \pm 0.077$, respectively. These values are in relatively good agreement with the photometrically-derived 2MASS $K_s$-band flux density ratio of $3.712 \pm 0.101$ relative to the PSF reference star.

After offsetting the POL*L-specific median-combined images by $\pm 1\sigma$ relative to their corresponding uncertainty image, we repeated the PSF subtraction procedure for each spacecraft orientation. We then calculated the uncertainty associated with each scale factor from the mean absolute difference between the upper and lower bound results, and estimate it to be $\lesssim 2\%$ for each PSF subtraction. Finally, we combined the resulting PSF scale factor uncertainties in quadrature with the POL*L-specific median-combined uncertainty images.

### 2.4. Linear Stokes Parameters

Adopting the matrix inversion method of Hines et al. (2000), we simultaneously derived the Stokes $I$, $Q$, and $U$ parameters using PSF-subtracted images from each spacecraft orientation. Elements of the coefficient matrix were assembled using polarizer properties determined from the post-NCS era (Batcheldor et al. 2009). Stokes images from each spacecraft orientation were rectified for geometrical distortion, rotated to a common orientation, smoothed using a one-resolution element ($\sim 0.''22$) Gaussian, and averaged to assemble the final images. Following the construction of the normalized Stokes parameters ($q = Q/I$ and $u = U/I$), we subsequently computed images of polarized intensity ($P = \sqrt{Q^2 + U^2}$), polarization fraction ($p = \sqrt{q^2 + u^2}$), and polarization position angle ($\psi = \frac{1}{2}\tan^{-1}(u/q)$, where $0° \leqslant \psi \leqslant 180°$) using standard formulae outlined in Tinbergen (2005).

Employing the prescription outline by Sparks & Axon (1999), we estimated uncertainties in the Stokes $I$, $Q$, and $U$ images using elements from the inverted coefficient matrix and the POL*L-specific PSF-subtracted uncertainty images for each spacecraft orientation. To construct the final uncertainty image for each Stokes parameter, we combined uncertainty images ($\sigma_i$) associated with each spacecraft orientation using $\sigma_{I,Q,U} = \sqrt{\sum_i^n \sigma_i^2}/\sqrt{n}$, where $n$ is the total number of images combined. Final uncertainties in the normalized Stokes parameters, polarized intensity, polarization fraction, and polarization position angle images were subsequently approximated following standard error propagation techniques.

The final Stokes images of the TW Hya disk are presented in Figure 1. The Stokes $I$, $Q$, and $U$ images are provided in the top panels, while the normalized Stokes images are included in the bottom panels for the sake of completeness. In addition, we present the final imaging polarimetry of the TW Hya disk in Figure 2. The polarized intensity and fractional polarization images are illustrated in the top panels, while the polarization position angle image is included in the lower panel for the sake of completeness. We note that the intensity images were converted to units of surface brightness using the POL0L photometric conversion factor of $5.994 \times 10^{-3}$ mJy s DN$^{-1}$.

### 2.5. Image Analysis

To estimate the surface density distribution of dust grains in the TW Hya disk, the intensity images were corrected for inverse-square illumination effects by the central star. Approximating the disk as an elliptical aperture, with a semi-minor-to-major axis ratio of 0.99 and a semi-major axis position angle of 155° (Andrews et al. 2016), we multiplied each pixel in the total and polarized intensity images by its projected squared pixel distance from the central star ($r^2$). Following the same procedure, we additionally constructed uncertainty images of the $r^2$-scaled intensities. The resulting $r^2$-scaled total and polarized intensity images are presented in Figures 1 and 2, respectively.

Prior to analyzing the $r^2$-scaled intensity and fractional polarization images, we constructed polar projections of each image by computing the average signal in concentric elliptical annuli from $\sim 0.''38$ ($\sim 23$ au) to $\sim 3.''34$ ($\sim 199$ au) at single pixel ($0.''07595$) intervals. In order to improve the signal-to-noise ratio at large projected distances from the central star, we divided each annulus into eight 45° wedge-shaped azimuthal bins. The signal uncertainty associated with each wedge was robustly calculated from the corresponding uncertainty image using $\sigma_s = \sqrt{\sum_i \sigma_i^2}/n_{\rm res}$, where $\sigma_i$ is the uncertainty measured in the $i$th pixel per wedge and $n_{\rm res}$ is the number of resolution elements per wedge. Since the width of a wedge-shaped aperture is approximately one-third the extent of a single resolution element, $n_{\rm res}$ was estimated using the number pixels along a single direction of the resolution element (i.e., $\sim 3$ pixels per resolution element, assuming a square-shaped aperture).

The polar projected $r^2$-scaled total and polarized intensity images are presented in Figures 3 and 5, respectively, while the polar projected fractional polarization image is presented in Figure 7. We de-biased the polarization fraction and polarized intensity polar projected images following the prescription of Plaszczynski et al. (2014). In particular, we used the modified asymptotic estimator method, in conjunction with the variance arithmetic mean, to compute de-biased point estimates of the polarized intensity and polarization fraction. We found that most of the resulting de-biased point estimates possess low signal-to-noise ratios ($S/N < 3.8$) and calculated their corresponding uncertainties using the 68% confidence interval expressions presented in Plaszczynski et al. (2014). We subsequently constructed the radial and azimuthal profiles of the total intensity, polarized intensity, and polarization fraction by averaging the polar projected images along the azimuthal-and radial-directions, respectively. For azimuthal angles containing less than three-fourths of the available pixels, no radially-averaged quantity was computed. Finally, we estimated the uncertainties associated with each profile following standard error propagation of the arithmetic mean.

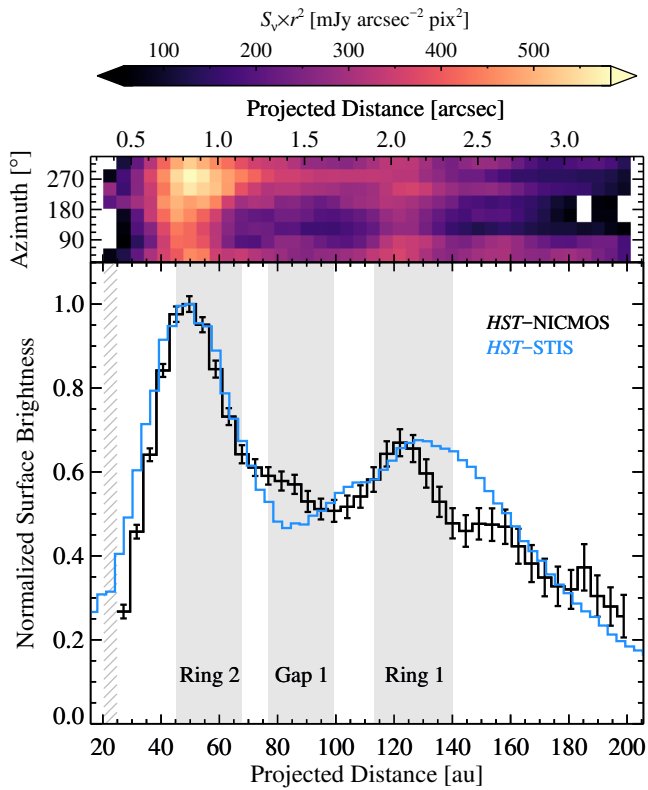

**Figure 3.** *HST*-NICMOS radial total intensity profile (black line) of the TW Hya disk. The normalized *HST*-STIS radial intensity profile (blue line) from Debes et al. (2017) is provided for comparison. Annuli coinciding with the ring 2 (∼45–68 au), gap 1 (∼77–99 au), and ring 1 (∼113–140 au) structures (solid gray regions), as well as regions of non-detection (diagonal gray lines), are indicated. The polar projection of the $r^2$-scaled total intensity image (top panel) is shown for reference, with missing data (white pixels) indicated.

## 3. RESULTS

### 3.1. *Total Scattered Light*

The $r^2$-scaled total intensity image of the TW Hya disk reveals a nearly face-on system with three distinct ring-shaped structures of varying intensity and width. These structures were previously detected in the *HST* optical and near-infrared scattered light observations by Debes et al. (2013). However, additional ring-shaped structures within the inner region ($r < 24$ au) of the TW Hya disk have since been discovered at optical (Debes et al. 2016, 2017), near-infrared (Akiyama et al. 2015; Rapson et al. 2015), and submillimeter (Andrews et al. 2016; Tsukagoshi et al. 2016) wavelengths. Adopting the universal nomenclature defined by van Boekel et al. (2017), we consequently refer to these bright and dark structures as ring 2, gap 1, and ring 1. The ring 2 structure extends from ∼0″.6 (∼36 au) to ∼1″.2 (∼71 au) and possesses a region of increased brightness at position angles between 225° ≲ $\phi$ ≲ 315° (east of north). The ring 1 structure possesses a similar width, extending from ∼1″.9 (∼113 au) to ∼2″.4 (∼143 au), but is considerably fainter than ring 2. Finally, situated between ring 2 and ring 1, the gap 1 structure exhibits a region of decreased surface brightness between 70° ≲ $\phi$ ≲ 200°.

#### 3.1.1. *Radial Surface Brightness*

The total intensity radial profile of the TW Hya disk is presented in Figure 3, compared to the averaged 2015-2016 *HST*-STIS profile from Debes et al. (2017). We detect the disk from ∼27 au to ∼199 au at signal-to-noise ratios between 5$\sigma$ and 55$\sigma$. However, we cannot constrain the disk total surface brightness near the inner working angle of ∼23 au. This result is likely a consequence of decreased surface brightness near the previously discovered ∼22 au gap (Akiyama et al. 2015; Rapson et al. 2015; Debes et al. 2017), rather than an instrumental effect near the coronagraphic obscuration. The surface brightness maxima of the ring 2 and ring 1 structures are found at ∼50 au and ∼122 au, respectively. Although these structures possess similar widths, the maximum surface brightness of ring 2 is a factor of 1.49 greater than that of ring 1. In addition, we find that the profile shape of the gap 1 structure exhibits a shallow slope that decreases by ∼12% from ∼77 au to ∼104 au.

The overall shape of the NICMOS radial surface brightness profile is generally consistent with that of the STIS profile. In particular, the peak position, width, and relative strength of ring 2 in the STIS profile is in excellent agreement with those found in the NICMOS profile. However, some differences between the surface brightness profiles are evident at projected distances greater than ∼73 au. Specifically, we find that the shape of the gap 1 structure in the STIS profile exhibits a minimum surface brightness at ∼83 au, and is substantially steeper than that found in the NICMOS profile. Moreover, the shape of ring 1 in the STIS profile is significantly broader than the NICMOS profile from ∼127 au to ∼154 au. Given the very good agreement between the NICMOS and STIS profiles at projected distances less than ∼127 au, the latter discrepancy is likely attributed to imperfect PSF subtraction resulting from the spectral type difference between TW Hya and BD +32°3739.

#### 3.1.2. *Azimuthal Surface Brightness*

To characterize the azimuthal brightness variations present in the $r^2$-scaled total intensity image, we gener-ated the azimuthal profile of the TW Hya disk at radii coinciding with the ring 2 (∼45–68 au), gap 1 (∼77–99 au), and ring 2 (∼113–140 au) structures. Following the method of Debes et al. (2013, 2017), we normalized each profile by its corresponding radial surface brightness to permit unbiased averaging within each wedge-shaped bin. The resulting azimuthal surface brightness profiles for the three annuli are presented in Figure 4. We find that the TW Hya disk exhibits a strong azimuthal brightness asymmetry at projected distances between ∼45 au and ∼99 au. At distances between ∼113 au and ∼140 au, the surface brightness asymmetry is considerably weaker but qualitatively similar to that found within the ring 2 and gap 1 structures. We note that the resulting profiles are not very sensitive to the precise centering of the azimuthal bins relative to the location of the central star (e.g., misalignments up to ∼$0\farcs1$).

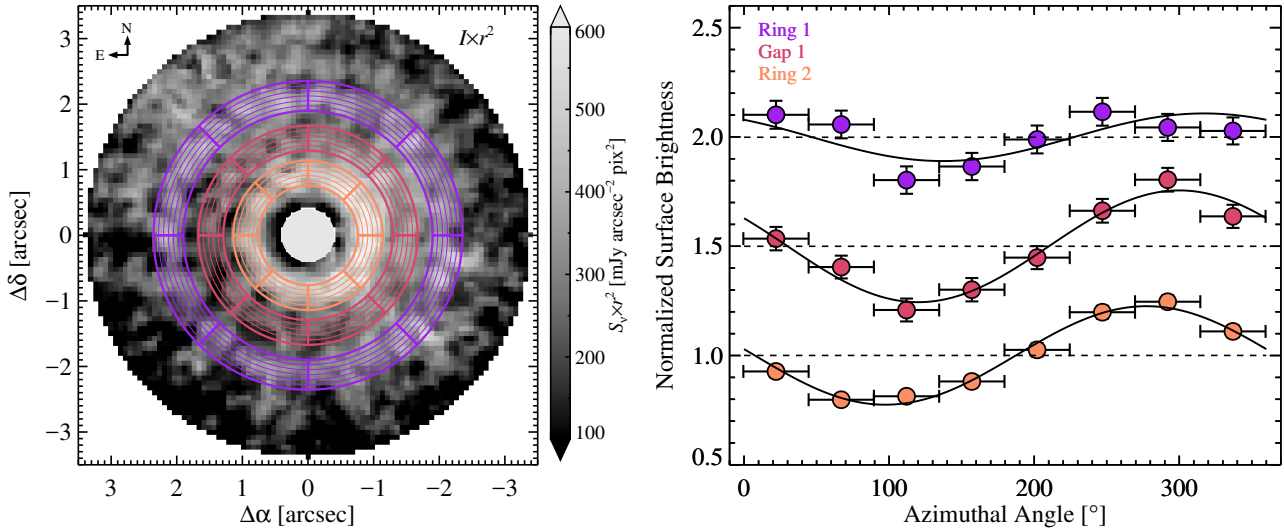

**Figure 4.** *HST*-NICMOS azimuthal total intensity profiles of the TW Hya disk (right panel). The profiles were constructed by averaging the surface brightness per wedge-shaped azimuthal bin along the ring 2 (orange circles; ∼45–68 au), gap 1 (red circles; ∼77–99 au), and ring 1 (violet circles; ∼113–140 au) structures; each profile was subsequently normalized by its corresponding azimuthally-averaged radial surface brightness. Azimuthal brightness asymmetries are found along ring 2, gap 1, and ring 1, as indicated by their best-fit sinusoidal model functions (black lines). The $r^2$-scaled total intensity image (left panel) is provided for reference, displaying the 45° wedge-shaped azimuthal bins. The image is displayed using an inverse hyperbolic sine stretch. A $0\farcs4$-radius coronagraphic mask (light gray circle) is indicated at the center of the image.

The TW Hya disk has long been known to possess an azimuthal surface brightness asymmetry at projected distances between ∼50 au and ∼140 au (Roberge et al. 2005; Debes et al. 2013). The origin of the surface brightness asymmetry has been attributed to the presence of strongly forward-scattering grains (Roberge et al. 2005), a shadow cast by a warped inner disk structure (Roberge et al. 2005; Rosenfeld et al. 2012), or inclination effects by a flared disk (Debes et al. 2013). In a recent multi-epoch study of the TW Hya disk, Debes et al. (2017) determined that the maximum brightness of the asymmetry is not stationary with time but rotates in the counterclockwise direction at a constant angular velocity of $22\fdg7$ yr$^{-1}$. The measured velocity between ∼50 au and ∼140 au greatly exceeds the expected Keplerian velocity at these projected distances by one to two orders of magnitude, suggesting that the asymmetry is a consequence of self-shadowing that occurs within the inner ($r < 5.6$ au) regions of the TW Hya disk. Debes et al. (2017) hypothesize that the shadow is produced by a warped or inclined inner ($r < 1$ au) disk structure that is precessing due to the presence of a 1.2 $M_{\rm Jup}$ planetary companion at 1.1 au.

Adopting an approach similar to that of Debes et al. (2017), the azimuthal surface brightness profile for each radius was modeled using a sinusoidal function:

$$f(\phi) = A\sin(\phi - B) + C, \qquad (1)$$

where $\phi$ is the position angle of the disk measured east of north, $A$ is the amplitude of the asymmetry, $B$ is the phase shift, and $C$ is the vertical offset. We determined the best-fit model parameters for each annulus using a Levenberg-Marquardt nonlinear least-squares minimization algorithm (Markwardt 2009). The formal uncertainties are based on propagated errors from the azimuthal surface brightness profiles. The best-fit model parameters and uncertainties for each annulus are reported in Table 2. The position angle of maximum surface brightness for each annulus was deduced from the best-fit phase shift parameter using $\phi_{\rm max} = B + (\pi/2)$. Similarly, the position angle of minimum surface brightness corresponding to each structure was inferred from $\phi_{\rm min} = \phi_{\rm max} - \pi$.

**Table 2.** Summary of Best-fit Sinusoidal Model Parameters

| Disk Structure | Projected Distance (au) | Amplitude $A$ | Phase Shift $B$ (°) | Vertical Offset $C$ | Position Angles[a] $\phi_{\max}$ (°) | $\phi_{\min}$ (°) | $\chi^2_\nu$ | $\nu$ |
|---|---|---|---|---|---|---|---|---|
| Total Intensity | | | | | | | | |
| ring 2 | 45–68 | 0.23 ± 0.01 | 187 ± 4 | 1.00 ± 0.01 | 277 ± 4 | 97 ± 4 | 0.77 | 5 |
| gap 1 | 77–99 | 0.26 ± 0.03 | 210 ± 6 | 1.00 ± 0.02 | 300 ± 6 | 120 ± 6 | 0.89 | 5 |
| ring 1 | 113–140 | 0.11 ± 0.03 | 227 ± 17 | 1.00 ± 0.02 | 317 ± 17 | 137 ± 17 | 1.96 | 5 |
| Polarized Intensity | | | | | | | | |
| ring 2 | 45–68 | 0.18 ± 0.05 | 183 ± 17 | 1.00 ± 0.04 | 273 ± 17 | 93 ± 17 | 0.54 | 5 |
| gap 1 | 77–99 | 0.21 ± 0.10 | 178 ± 28 | 1.00 ± 0.07 | 268 ± 28 | 88 ± 28 | 0.14 | 5 |
| Non-Polarized Intensity | | | | | | | | |
| ring 2 + gap 1 | 45–99 | 0.21 ± 0.06 | 204 ± 17 | 1.00 ± 0.04 | 294 ± 17 | 114 ± 17 | 0.27 | 5 |

NOTE—Formal uncertainties are based on propagated errors from the radially-averaged azimuthal surface brightness profiles.

[a] Position angles of maximum and minimum surface brightness measured east of north. The position angle of maximum brightness was derived from the best-fit phase shift parameter using $\phi_{\max} = B + (\pi/2)$, while the position angle of minimum brightness was calculated by $\phi_{\min} = \phi_{\max} - \pi$.

A comparison between the best-fit sinusoidal models and the observed azimuthal surface brightness profile for each annulus is provided in Figure 4. In accord with Debes et al. (2017), we find that the brightness asymmetries associated with the ring 2 and gap 1 structures can be adequately ($\chi^2_\nu \approx 0.8$–0.9) simulated using a sinusoidal function. However, the asymmetry present along ring 1 is less well-constrained ($\chi^2_\nu \approx 2$) and possesses an amplitude that is ∼42%–∼48% weaker in comparison. The ratio of maximum-to-minimum surface brightness along the ring 2 and gap 1 structures indicates that the north-west region of the disk is a factor of ∼1.6–1.7 brighter than the south-east region. The position angle of maximum brightness along these annuli is found at $277 \pm 4°$ and $300 \pm 6°$, respectively. These results are in relatively good agreement with the position angle of maximum surface brightness predicted by Debes et al. (2017) for the 2004 epoch, suggesting that the azimuthal brightness variations observed along the ring 2 and gap 1 structures are a consequence of self-shadowing effects.

### 3.2. *Polarized Scattered Light*

The $r^2$-scaled polarized intensity image of the TW Hya disk is presented in Figure 2. While the overall surface brightness distribution of the disk is similar to that found in total intensity, the brightness contrast between the ring and gap structures is considerably lower in polarized intensity. The ring 2, gap 1, and ring 1 disk structures have been recently detected in polarized scattered light by van Boekel et al. (2017), using the Spectro-Polarimetric High-contrast Exoplanet REsearch (SPHERE) instrument on the Very Large Telescope (VLT). Therefore, a comparison between the NICMOS and SPHERE surface brightness profiles may be informative, particularly in regard to self-shadowing effects in the polarized intensity of the TW Hya disk.

#### 3.2.1. *Radial Surface Brightness*

The polarized intensity radial profile of the TW Hya disk is presented in Figure 5, compared to the VLT-SPHERE $H$-band profile from van Boekel et al. (2017). We detect the disk from the inner working angle of ∼23 au to ∼163 au at signal-to-noise ratios between $3\sigma$ and $14\sigma$. However, beyond the ring 1 structure, the signal-to-noise ratio of the disk is too low to reliably constrain the shape of the polarized intensity profile. The surface brightness maxima of the ring 2 and ring 1 structures are found at ∼50 au and ∼117–127 au, respectively. However, the relative surface brightness of ring 2 is a factor of 1.54 greater than that of ring 1. Although these results are generally consistent with those from our total intensity analysis, we find that the widths of the ring 2 and ring 1 structures are substantially broader in the polarized intensity profile. Moreover, the profile shape of the gap 1 annulus exhibits a shallow slope that increases by ∼6% from ∼90 au to ∼104 au. Finally, we find that the surface brightness ratio of the ring 2 and ring 1 structures relative to gap 1 is 1.72 and 1.12, respectively.

Given the uncertainties associated with the NICMOS polarized surface brightness, the general shape of the radial profile agrees reasonably well with that of the VLT-SPHERE $H$-band profile. Specifically, the peak position and inner edge of the ring 2 structure is in good agreement with that found in the SPHERE profile. However, differences between the profiles are apparent beyond ∼50 au. In particular, we find that the outer edge of ring 2 is noticeably narrower from ∼59 au to ∼77 au in the SPHERE profile. Furthermore, the polarized surface brightness ratio of ring 1 relative to gap 1 is a factor of ∼2 larger in the SPHERE profile. Although these discrepancies are likely attributed to imperfect PSF subtraction resulting from dissimilar spectral types, we note that a factor of ∼4 difference in spatial resolution between the SPHERE and NICMOS observations may give rise to the broadening and shallowing of structures found in the NICMOS polarized intensity profile.

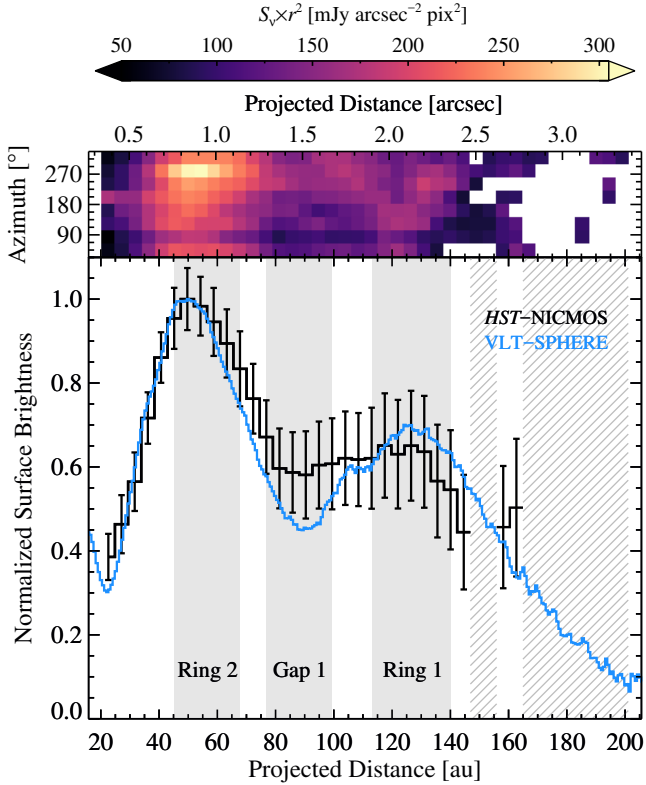

**Figure 5.** *HST*-NICMOS radial polarized intensity profile (black line) of the TW Hya disk. The normalized VLT-SPHERE radial polarized intensity profile (blue line; *H*-band) from van Boekel et al. (2017) is provided for comparison. Annuli coinciding with the ring 2 (∼45–68 au), gap 1 (∼77–99 au), and ring 1 (∼113–140 au) structures (solid gray regions), as well as regions of low signal-to-noise ratio ($P/\sigma_P < 1$; diagonal gray lines), are indicated. The polar projection of the $r^2$-scaled polarized intensity image (top panel) is shown for reference, with missing data (white pixels) indicated.

### 3.2.2. *Azimuthal Surface Brightness*

In order to explore the significance of the azimuthal brightness variations observed in the $r^2$-scaled polarized intensity image, the azimuthal profile of the TW Hya disk was similarly computed along the ring 2, gap 1, and ring 1 structures. The azimuthal surface brightness profile for each annulus is presented in Figure 6 and reveals a significant asymmetry along the ring 2 and gap 1 structures. The overall appearance of the azimuthal asymmetry is qualitatively similar to that found in the total intensity profile. However, no significant asymmetry is found along the ring 1 annulus in polarized scattered light. It is important to note that the signal-to-noise ratio per azimuthal bin at the ring 2 annulus is a factor of ∼2.8 greater than that found at the ring 1 annulus.

The azimuthal surface brightness profiles of the ring 2 and gap 1 structures were similarly modeled using the sinusoidal function described in Equation 1. The best-fit model results are presented in Table 2. The brightness variations along the ring 2 and gap 1 annuli can be adequately ($\chi_\nu^2 = 0.54$ and $0.14$, respectively) simulated using a sinusoidal model. We find that the best-fit model parameters for the ring 2 annulus are nearly identical to those found for the gap 1 annulus. The ratio of maximum-to-minimum surface brightness along these annuli suggests that the north-west region of the disk is a factor of ∼1.4–1.5 brighter than the south-east region. Furthermore, the position angle of maximum surface brightness along the ring 2 and gap 1 annuli is found at $273 \pm 17°$ and $268 \pm 28°$, respectively. We find that these results are consistent with those from our total intensity analysis, indicating that the polarized scattered light from the TW Hya disk is also affected by self-shadowing effects.

A comparison between the NICMOS and SPHERE surface brightness asymmetries reveals unexpected differences between the profiles. In particular, the SPHERE surface brightness profiles do not possess a single maximum, but instead exhibit a set of weaker maxima at 125° and 225° for the ring 2 annulus, 125° and 255° for the gap 1 annulus, and 135° and 225° for the ring 1 annulus. Furthermore, the brightness variation along the ring 1 annulus is qualitatively similar to those found for the ring 2 and gap 1 annuli. In fact, van Boekel et al. (2017) found that the shape of the SPHERE azimuthal profile is remarkably consistent from ∼13 au to ∼143 au. We note this result is incompatible with the results of Debes et al. (2017), who found that azimuthal brightness variations interior to 50 au are fundamentally different than those at larger projected distances and thus unaffected by self-shadowing.

While quantifying these differences with great certainty is beyond the scope of this study, we propose that temporal changes in the azimuthal surface brightness may be a consequence of inclination-induced polarization effects. For example, the azimuthal polarized intensity profiles of face-on ($i < 15°$) disks are expected to exhibit sinusoidal-like behavior, with regions

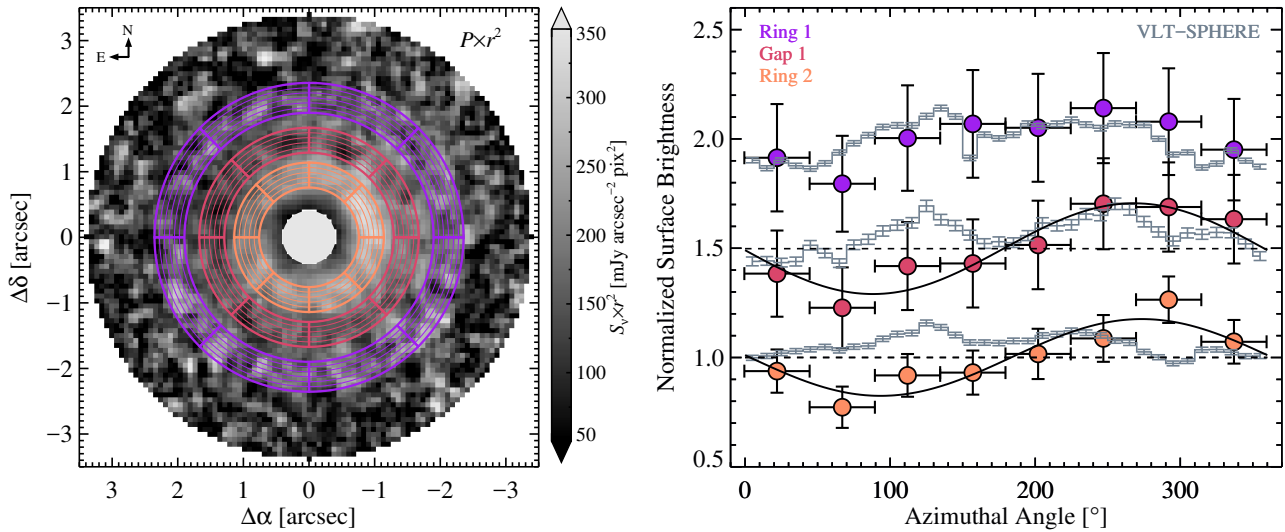

**Figure 6.** *HST*-NICMOS azimuthal polarized intensity profile of the TW Hya disk (right panel). The profiles were constructed by averaging the surface brightness per wedge-shaped azimuthal bin along the ring 2 (orange circles; ∼45–68 au), gap 1 (red circles; ∼77–99 au), and ring 1 (violet circles; ∼113–140 au) structures; each profile was subsequently normalized by its corresponding azimuthally-averaged radial surface brightness. Azimuthal brightness asymmetries are found along ring 2 and gap 1, as indicated by their best-fit sinusoidal model functions (black lines). The VLT-SPHERE azimuthal polarized intensity profiles (dark gray lines; *H*-band) from van Boekel et al. (2017) are shown for comparison; each profile was normalized and shifted for clarity. The $r^2$-scaled polarized intensity image (left panel) is provided for reference, displaying the 45° wedge-shaped azimuthal bins. The image is displayed using an inverse hyperbolic sine stretch. A $0\rlap{.}''4$ radius coronagraphic mask (light gray circle) is indicated at the center of the image

of minimum surface brightness occurring along the disk semi-minor axes (Jang-Condell 2017); similar geometrical effects have also been detected in more inclined disk systems (e.g., Oppenheimer et al. 2008; Perrin et al. 2009). If this time-invariant, sinusoidal component cancels out the self-shadowing sinusoidal component in the SPHERE azimuthal profiles, then the observed differences between the NICMOS and SPHERE profiles may be reconciled. In addition, the observed differences between the azimuthal brightness profiles may also be a consequence of multiple scattering in the TW Hya disk. We note that multiple scattering in optically thick protoplanetary disks may result in non-azimuthal polarization, which is not well-quantified by the radial Stokes formalism often employed in ground-based polarimetric studies (Canovas et al. 2015).

### 3.3. *Linear Polarization*

The polarization fraction of the TW Hya disk is presented in Figure 2 and reveals a centro-symmetric pattern of polarization vectors surrounding the central star. Despite being the target of numerous ground-based polarimetric studies, relatively little insight exists into the polarization fraction of the TW Hya disk. Qualitative comparisons between ground-based polarized intensity profiles and space-based total intensity profiles first suggested that the polarization fraction of the disk is nearly constant with radial distance (e.g., Apai et al. 2004). Follow-up measurements by Hales et al. (2006), using UKIRT-IRPOL2 *J*- and *H*-band imaging polarimetry and *HST*-NICMOS total intensity coronagraphic imaging from Weinberger et al. (2002), predict a maximum polarization fraction of ∼0.30 at ∼70 au. More recently, van Boekel et al. (2017) estimate a disk polarization fraction of ∼0.35 by comparing VLT-SPHERE *H*-band observations with *HST*-NICMOS F160W observations from Debes et al. (2016). However, a more reliable (and direct) method in determining the polarization fraction of the TW Hya disk requires simultaneously-derived Stokes *I*, *Q* and *U* parameters from contemporaneous observations.

#### 3.3.1. *Radial Polarization Fraction*

The radial polarization fraction profile of TW Hya is presented in Figure 7 and reveals that the disk is adequately ($3 \lesssim p/\sigma_p \lesssim 13$) detected from ∼32 au to ∼136 au. However, between ∼140 au and ∼199 au, the polarization fraction of the disk is not reliably recovered due to the low signal-to-noise ratio ($p/\sigma_p < 1$) present at large projected radii. Contrary to that predicted by ground-based polarimetric studies, we find that the polarization fraction of the TW Hya disk is not constant with distance from the central star. In particular, the fractional polarization profile varies from $0.63 \pm 0.09$ to $0.46 \pm 0.10$ and exhibits three trough-like features that coincide with the ring 2, gap 1, and ring 1 structures. The polarization fraction along these structures decreases by a maximum of ∼24%, ∼20%, and ∼17%, respectively. We note that a region of lower polarization at optical wavelengths was also predicted for ring 2 by van Boekel et al. (2017), who qualitatively compared the VLT-SPHERE $R'$- and $I'$-band polarized intensity profiles to the STIS total intensity profile from Debes et al. (2016). The factor of ∼2 discrepancy between the polarization fraction measured from this study and those estimated from previous ground-based studies is likely the result of systematic errors arising from the combination of non-contemporaneous observations from ground- and space-based facilities (e.g., Quanz et al. 2011).

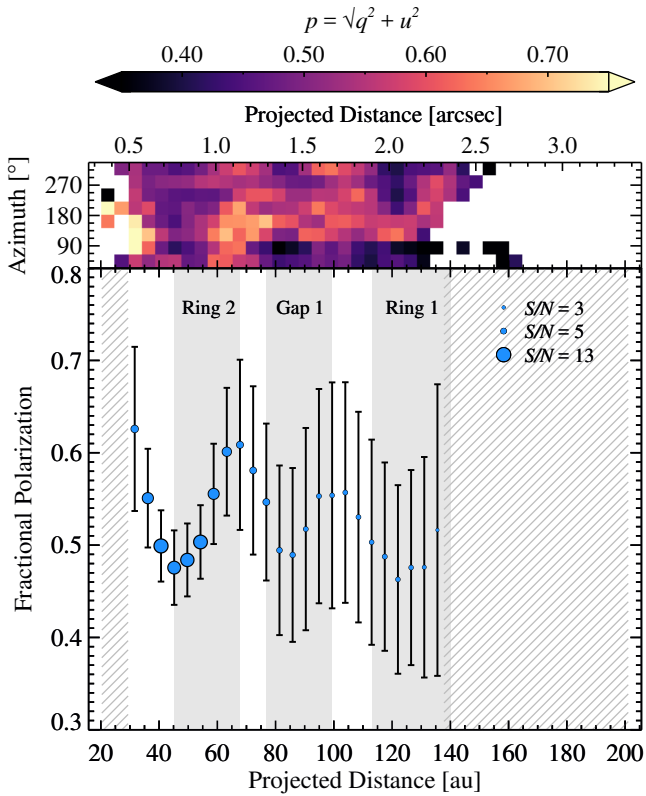

**Figure 7.** *HST*-NICMOS radial polarization fraction profile (blue circles) of the TW Hya disk. The magnitude of each data point represents the final signal-to-noise ratio of the fractional polarization. Annuli coinciding with the ring 2 (∼45–68 au), gap 1 (∼77–99 au), and ring 1 (∼113–140 au) structures (solid gray regions), as well as regions of low signal-to-noise ratio ($p/\sigma_p < 1$; diagonal gray lines), are indicated. The polar projection image of the polarization fraction is shown for reference (top panel), with missing data (white pixels) indicated.

The observed radial variation in the polarization fraction is in agreement with the scattered light models by Jang-Condell (2017), who found that regions of decreased polarization fraction coincide with surface brightness maxima in face-on protoplanetary disks with planet-induced gaps. This phenomenon may be explained by the relative proportion of multiple-to-single scattering in the total scattered light intensity. For gaps created by 70 and 200 $M_\oplus$ planets, Jang-Condell (2017) predicts polarization fraction variations of ∼10% and ∼15%, respectively, with a greater proportion of multiple scattering at projected radii immediately outside a planet-induced gap. In addition, we note that multiple scattering effects are also proposed to reduce the fractional polarization in low surface density regions of partially filled gaps (van Boekel et al. 2017). In this scenario, the polarized intensity from this region becomes dominated by photons that experience a minimum of two scattering events; first into the gap region and then scattered into the line of sight.

### 3.3.2. *Azimuthal Polarization Fraction*

Having found sinusoidal brightness asymmetries in the total and polarized intensity profiles, we finally explore the possible presence of azimuthal variations in the polarization fraction of the TW Hya disk. The azimuthal profile, presented in Figure 8, was constructed by averaging the fractional polarization over annuli coinciding with the ring 2 and gap 1 structures (∼45–99 au). We find that the polarization fraction of the disk is not constant with position angle, but possesses a strong variation at position angles coinciding with the shadowed region ($90° \lesssim \phi \lesssim 225°$). In particular, the azimuthal profile exhibits a factor of ∼1.2 increase in polarization fraction between ∼90° and ∼180°, followed by a more gradual decline from ∼180° to ∼315°. Moreover, the polarization fraction in the west-south-west region of the TW Hya disk is a factor of ∼1.10 greater than that in the east-north-east region. Since backscattering is expected to increase the polarization fraction on the far side of a disk composed of small scattering grains (Jang-Condell 2017), this result suggests that the east-north-east region ($\phi \approx 65°$) is associated with the near side of the disk. We note that this finding is in contrast with the prediction by van Boekel et al. (2017), who found that the west-south-west region ($\phi \approx 245°$) was brighter in polarized intensity and considered it to be the near side of the disk.

## 4. DISCUSSION

Detecting an increase in the fractional polarization along the shadowed region of a protoplanetary disk may be a consequence of geometrical effects (e.g., inclination, disk flaring) or a reduction in the relative proportion of multiple-to-single scattering in the total scattered light intensity. In order to explore whether the observed behavior is a result of geometrical effects, Rayleigh scattering models of a nearly face-on disk were constructed following an approach similar to that described in Tanii et al. (2012). Assuming that the scattering surface of the disk is predominantly comprised of grains in the Rayleigh limit ($\lambda \gg 2\pi a$, where $a$ is the radius of the largest particles), the polarization fraction

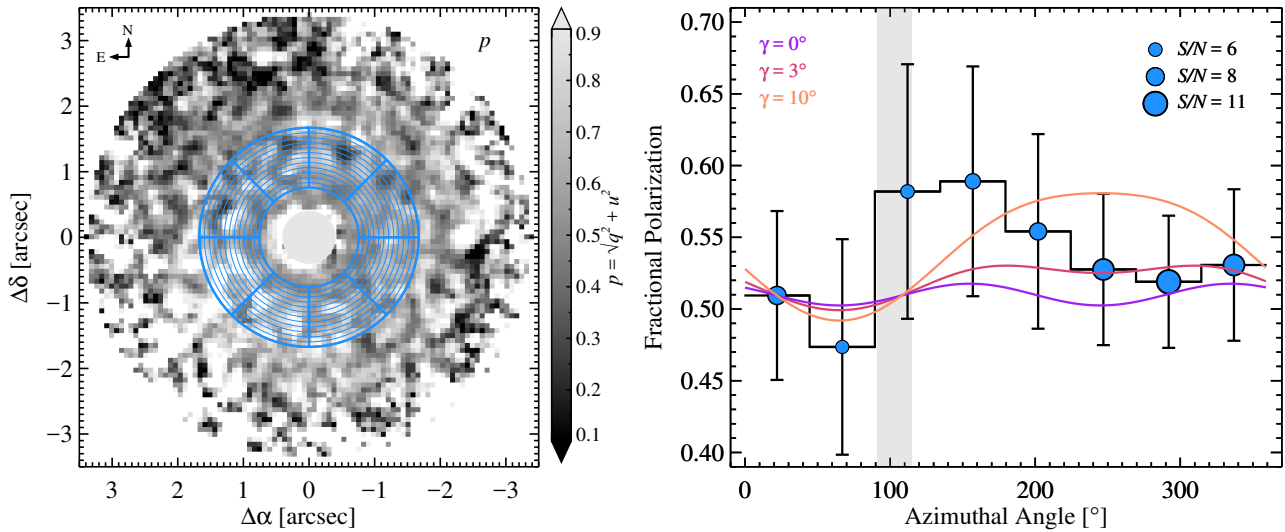

**Figure 8.** *HST*-NICMOS azimuthal polarization fraction profile of the TW Hya disk (right panel). The profile (blue circles) was constructed by averaging the fractional polarization per wedge-shaped azimuthal bin over the ring 2 and gap 1 (∼45–99 au) structures. The magnitude of each data point represents the final signal-to-noise ratio of the polarization fraction. For comparison, Rayleigh scattering models are illustrated for a nearly face-on disk ($i = 7°$) with different degrees of flaring: $\gamma = 0°$ (violet line), $\gamma = 3°$ (red line), and $\gamma = 10°$ (orange line). The models were normalized and subsequently scaled to match the observed polarization fraction profile at 22°. The position angle of minimum total scattered light intensity (gray region), deduced from the best-fit sinusoidal model results of ring 1, is indicated. The fractional polarization image (left panel) is presented for reference, displaying the 45° wedge-shaped azimuthal bins. The image is displayed using a inverse hyperbolic sine stretch. A 0″.4 radius coronagraphic mask (light gray circle) at the center of the image and missing data (white pixels) are indicated.

of single scattered light was approximated using

$$p = \frac{1 - \cos^2(\theta)}{1 + \cos^2(\theta)}. \quad (2)$$

In the context of our *HST*-NICMOS observations, we note that this scattering approximation is valid for grains with radii much smaller than $a \approx 0.3$ μm.

Adopting the flared disk model geometry derived by McCabe et al. (2002), the scattering angle ($\theta$) is related to the position angle of the disk ($\phi$), the position angle corresponding to the near side of the disk semi-minor axis ($\phi_0 = 65°$), and the disk inclination angle ($i = 7°$) by

$$\cos(\theta + \gamma) = \sqrt{1 - \frac{1}{1 + \cos^2(\phi - \phi_0)\tan^2(i)}}(-1)^j, \quad (3)$$

where $j = 1$ for $\cos(\phi - \phi_0) < 0$ and $j = 0$ for $\cos(\phi - \phi_0) > 0$. The disk flaring angle ($\gamma$) is given by $\gamma = \tan^{-1}(h/r)$, where $h$ is the vertical height of the disk scattering surface ($\tau = 2/3$) and $r$ is the projected mid-plane radius. Based on the radiative transfer model results from van Boekel et al. (2017), the $H$-band aspect ratio of the TW Hya disk is $h/r \approx 0.18$ at 100 au. Therefore, we consider only disk flaring angles up to $\gamma = 10°$.

The resulting Rayleigh scattering models are provided in Figure 8. To facilitate a meaningful comparison with the observed profile, the models were normalized and subsequently scaled to match the polarization fraction at 22°. We find that the polarization fraction along the unshadowed region ($225° \lesssim \phi \lesssim 90°$) of the TW Hya disk can be approximated by a moderately flared disk composed of sub-micron-sized scattering grains. More specifically, a disk flaring angle of at least $\gamma = 3°$ is required to replicate the slightly higher polarization fraction observed along the far side of the disk. However, upon further inspection, we find that the moderately flared disk model is unable to simulate the higher polarization fraction measured along the shadowed region ($90° \lesssim \phi \lesssim 225°$) of the disk. Increasing the disk flaring angle to $\gamma = 10°$ results in an overestimate of the fractional polarization along the far side of the disk semi-minor axis ($\phi \approx 245°$) and virtually no change along the shadowed region ($\phi \approx 112°$). In light of these findings, we conclude that the observed increase in polarization fraction along the shadowed region is not the result of geometrical effects arising from disk flaring.

To further investigate whether the observed variation in polarization fraction is caused by a reduction in the relative proportion of multiple-to-single scattering in the total scattered light intensity, we calculated the non-polarized intensity of the TW Hya disk by subtracting the polarized intensity from the total intensity. For reference, the resulting image is presented in the middle panel of Figure 1 and traces the multiple scattering component across the disk surface. However, because the non-polarized intensity is derived from the polarized intensity, this quantity must also be de-biased before constructing its surface brightness distribution. Therefore, the de-biased non-polarized intensity was calculated by subtracting the polar projected (de-biased) polarized intensity image from the polar projected total intensity image. The azimuthal profile of the non-polarized intensity was subsequently computed by averaging the surface brightness of the disk over annuli coinciding with the ring 2 and gap 1 structures (∼45–99 au). A comparison between the non-polarized and polarized intensity profiles is provided in Figure 9. The polarized intensity profile was normalized and scaled to match the maximum surface brightness of the non-polarized scattered light. Similar to the total and polarized intensity profiles, the non-polarized intensity profile exhibits strong sinusoidal-like behavior that is the result of self-shadowing. Utilizing the sinusoidal model function described in Equation 1, we find that the amplitude of the non-polarized intensity profile is factor of 1.15 stronger than that found in the polarized intensity profile. These results indicate that a reduction in the relative proportion of multiple-to-single scattering in the total scattered intensity is present in the shadowed region of the TW Hya disk.

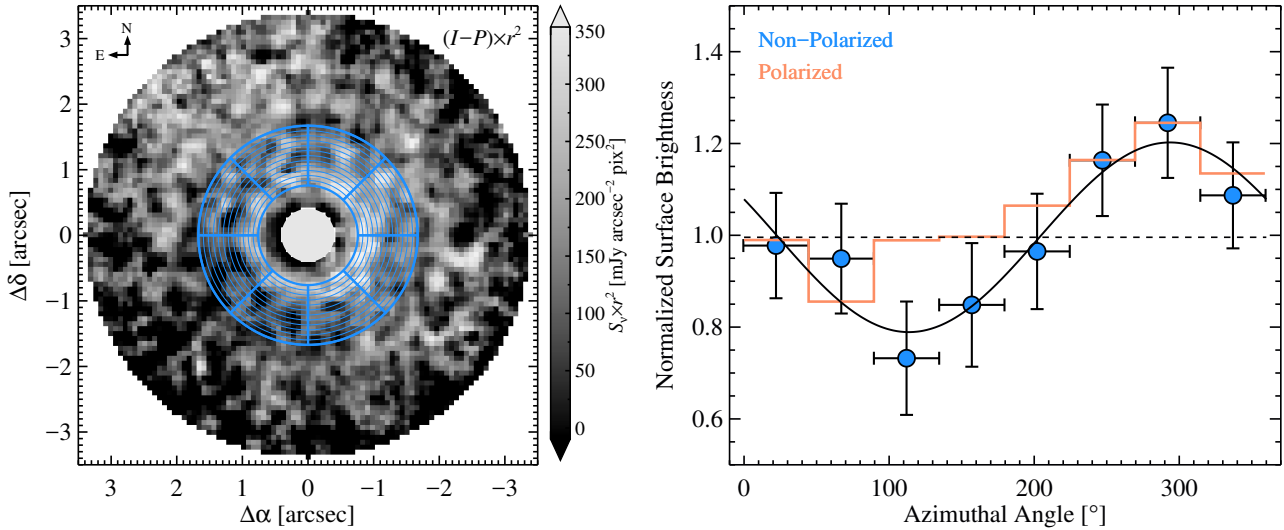

**Figure 9.** Comparison between the *HST*-NICMOS azimuthal non-polarized (blue circles) and polarized (orange line) intensity profiles of the TW Hya disk (right panel). The non-polarized intensity profile was constructed by averaging the surface brightness per wedge-shaped azimuthal bin over the ring 2 and gap 1 (∼45–99 au) structures. A strong brightness asymmetry is present along the ring 2 and gap 1 structures, as indicated by the best-fit sinusoidal model function (black line). The polarized intensity profile, computed over the same radial extent, has been normalized and shifted to match the peak surface brightness of the non-polarized profile. The $r^2$-scaled non-polarized intensity image (left panel) is provided for reference, displaying the 45° wedge-shaped azimuthal bins. The image is displayed using an inverse hyperbolic sine stretch. A 0.″4 radius coronagraphic mask (light gray circle) is indicated at the center of the image

Recent ground-based polarimetric observations of the SR 21 (Follette et al. 2013) and RY Tau (Takami et al. 2013) disks suggest that the small (sub-micron-sized) grain scattering surfaces of protoplanetary disks may be comprised of vertically stratified layers of different optical thickness. In particular, Follette et al. (2013) and Takami et al. (2013) propose a two-layered scattering surface consisting of an overlying optically thin component and underlying optically thick component. In the case of SR 21, Follette et al. (2013) propose that polarized scattered light from the disk is dominated by the overlying optically thin component, which contributes ∼30 times more flux than the underlying optically thick component. Although single scattering is also considered to be the primary mechanism of polarization in the overlying optically thin component of the RY Tau disk, Takami et al. (2013) hypothesize that half of the photons in this layer are scattered toward the underlying optically thick component. This scenario implies that light scattered from the underlying optically thick component has undergone multiple scattering events.

Employing a scattering structure analogous to that described for the SR 21 and RY Tau disks, we propose that the observed polarization fraction of the TW Hya disk may result from the combination of (1) single scattering in an overlying optically thin layer and (2) multiple scattering in an underlying optically thick layer. However, to account for the higher polarization fraction measured along the shadowed region ($90° \lesssim \phi \lesssim 225°$) of the disk, the contribution of single scattering in the overlying optically thin layer must be greater than the contribution of multiple scattering in the underlying optically thick layer. The diminished contribution of multiple scattering in the optically thick layer may naturally result from a shadow cast by an inclined inner disk, which prevents direct stellar light from reaching the underlying scattering surface of the disk at projected radii falling below some vertical surface height.

## 5. CONCLUSIONS

We have presented *HST*-NICMOS coronagraphic imaging polarimetry of the protoplanetary disk surrounding TW Hya. In contrast to previous ground-based polarimetric studies, these observations simultaneously reveal the nearly face-on disk in both total and polarized intensity. We reliably detect the TW Hya disk on angular scales between ∼27 au and ∼199 au (∼23 au and ∼163 au) in total (polarized) scattered light and resolve the previously reported ring-shaped structures: ring 2, gap 1, and ring 1. Furthermore, we find that the azimuthal surface brightness distribution of the disk is highly asymmetric in both total and polarized scattered light. In particular, strong sinusoidal-like variations are present along the ring 2 (∼45–68 au) and gap 1 (∼77–99 au) structures, with the maximum surface brightness of the disk occurring at position angles between ∼268° and ∼300°. However, the brightness variations in total intensity are ∼24%–28% stronger than those in polarized intensity. Our results are consistent with those from the multi-epoch *HST* study by Debes et al. (2017), suggesting that the azimuthal brightness variations observed along the ring 2 and gap 1 structures are a consequence of self-shadowing effects.

In addition to the scattered light intensities, the polarimetric observations allow direct access to the spatial distribution of the polarization fraction across the surface of the TW Hya disk. Contrary to predictions by previous ground-based polarimetric studies, we find that the polarization fraction of the disk strongly varies along both the radial- and azimuthal-direction. In particular, radial variations of ∼24%, ∼20%, and ∼17% are present along the ring 2, gap 1, and ring 1 structures, respectively. In accord with Jang-Condell (2017), we find that these regions of lower polarization fraction are associated with annuli of increased surface brightness, suggesting that the relative proportion of multiple-to-single scattering is greater along the ring and gap structures. Moreover, averaging the polarization fraction over the ring 2 and gap 1 structures (∼45–99 au), we find that strong (∼20%) azimuthal variation is present along the shadowed region ($90° \lesssim \phi \lesssim 225°$) of the disk. Further investigation reveals that the azimuthal variation is not the result of disk flaring effects, but rather from a decrease in the relative contribution of multiple-to-single scattering along the shadowed region.

We propose that the azimuthal variations in the polarization fraction may result from a combination of (1) single scattering in a vertically-extended, overlying optically thin layer and (2) multiple scattering in an underlying optically thick layer. However, to account for the higher polarization fraction along the shadowed region of the disk, the contribution of non-polarized scattered light must be less than the contribution of polarized scattered light. The diminished contribution of multiple scattering in the optically thick component may be a natural consequence of shadowing by an inclined inner disk, which prevents direct stellar light from reaching the underlying surface layer of the disk.

Shadows cast on the outer disk by a warped or more inclined structure bears consequence on the relative orientation of the inner disk. In particular, the angular separation of the shadows cast on the outer disk is dependent on the relative inclination of the inner disk and the aspect ratio of the outer disk surface height (Min et al. 2017). Small differences in the relative inclination are expected to generate a single azimuthally-wide shadow, while larger differences are predicted to create two azimuthally-narrow shadows (e.g., Marino et al. 2015; Benisty et al. 2017). Following the analytical framework of Min et al. (2017), the relative difference between the inclination of the inner and outer disk may be directly inferred from

$$\tan(\Delta i) = \sqrt{\left(\frac{h}{r}\right)^2 \times \left(\tan^2\left(\frac{\omega}{2}\right) + 1\right)}, \qquad (4)$$

where $h/r$ is the aspect ratio of the disk and $\omega$ is the angular separation between shadows as measured in the plane of the outer disk. Adopting a disk aspect ratio of 0.14 at 45 au and 0.18 at 99 au from van Boekel et al. (2017), we find that the relative difference between the inclination of the inner and outer disk of TW Hya ranges from ∼8° to ∼11°. Although the sinusoidal-like brightness variations in the TW Hya disk are indicative of a single azimuthally-wide shadow, we note this calculation assumes that the angular separation of the shadows cast on the outer disk is less than 45° (i.e., the angular extent in which the surface brightness is measured). Placing this result in the context of the outer disk ($i = 7°$), we hypothesize that the inclination of the inner disk is between ∼15° and ∼18°, which is in accord with that predicted by Debes et al. (2017).

While resolving the inner 1 au region of the TW Hya disk is not presently possible with current space- and ground-based coronagraphic, future observations with the next generation of ground-based telescopes may be capable of accomplishing such a feat. In the meantime, detailed properties of the innermost regions of disks may be inferred through the study of shadows cast on their outer disk using high-contrast scattered light imaging from current space- and ground-based observatories and forthcoming coronagraphic observations with the *James Webb Space Telescope*.

(STScI/NASA), the Space Telescope European Coordinating Facility (ST-ECF/ESA), and the Canadian Astronomy Data Centre (CADC/NRC/CSA). Support for program number HST-AR-12630 was provided by NASA through a grant from the Space Telescope Science Institute, which is operated by the Association of Universities for Research in Astronomy, Incorporated, under NASA contract NAS 5-26555. The authors are grateful to Roy van Boekel for kindly providing the VLT-SPHERE polarized intensity profiles of the TW Hya disk and to the anonymous referee for thoughtful comments that improved the manuscript. C.A.P. thanks Stéphane Plaszczynski for insightful discussions regarding polarization de-biasing techniques.